\documentclass[%
 reprint,
 /amsmath,amssymb,
 aps,
]{revtex4-1}

\usepackage{epstopdf}
\usepackage[dvipdfmx]{graphicx}% Include figure files
\usepackage{color}
\usepackage{dcolumn}% Align table columns on decimal point
\usepackage{amsmath}
\usepackage{bm}% bold math
\begin{document}

\preprint{APS/123-QED}

\title{Highly Viscous Microjet Generator}% Force line breaks with \\

\author{Hajime Onuki}
\author{Yuto Oi}%
\author{Yoshiyuki Tagawa}%
 \email{tagawayo@cc.tuat.ac.jp} 
\affiliation{%
Department of Mechanical Systems Engineering, Tokyo University of Agriculture and Technology,
Naka-cho 2-24-16, Koganei, Tokyo 184-8588, Japan
}%

\date{\today}% It is always \today, today,

\begin{abstract}
This paper describes a simple yet novel system for generating a highly viscous microjet.
The jet is produced inside a wettable thin tube partially submerged in a liquid.
The gas-liquid interface inside the tube, which is initially concave,  is kept much deeper than that outside the tube.
An impulsive force applied at the bottom of a liquid container leads to significant acceleration of the liquid inside the tube followed by flow-focusing due to the concave interface.
The jet generation process can be divided into two parts that occur in different time scales, i.e. {\it the Impact time} (impact duration $\le O(10^{-4})$ s) and {\it Focusing time} (focusing duration $\gg O(10^{-4})$ s).
In Impact time, the liquid accelerates suddenly due to the impact.
In Focusing time, the microjet emerges due to flow-focusing.
In order to explain the sudden acceleration inside the tube in Impact time, we develop a physical model based on a pressure impulse approach.
Numerical simulations confirm the proposed model, indicating that the basic mechanism of the acceleration of the liquid due to the impulsive force is elucidated.
Remarkably, the viscous effect is negligible in Impact time.
In contrast, in Focusing time, the viscosity plays an important role in the microjet generation.
We experimentally and numerically investigate the velocity of microjets with various viscosities.
We find that higher viscosities lead to reduction of the jet velocity, which can be described by using Reynolds number (the ratio between the inertia force and the viscous force).
This novel device may be a starting point for next-generation technologies, such as high-viscosity inkjet printers including bioprinters and needle-free injection devices for minimally invasive medical treatments.
\end{abstract}

\pacs{Valid PACS appear here}% PACS, the Physics and Astronomy
\maketitle

%\tableofcontents

%Introduction
\section{\label{sec:1}INTRODUCTION}
Technologies for generating liquid jets \cite{jet} are utilized in important modern devices such as inkjet printers \cite{inkjet,inkjet2,inkjet3}.
However, most existing printers can eject only low-viscosity liquids of up to 20 mm$^2$/s, which is about twenty times the viscosity of water.
This limitation causes serious problems such as blurring and color-dulling.
In order to solve these problems, a method for generating jets of highly viscous liquids \cite{Zhang,viscous_patent} is desired.
Furthermore, the generation of viscous microjets can open a new door for next-generation technologies such as  needle-free injection devices \cite{needle-free1,needle-free2,needle-free3}, printed electronics \cite{Metal-printing}, 3D-printers \cite{3D-printing}, and bio-printers \cite{Bioprinting}, since many highly viscous liquids have various functionalities such as adherence and conductivity \cite{Func-drop}, which most low-viscosity liquids lack.

\begin{figure}
\includegraphics[scale = 0.13]{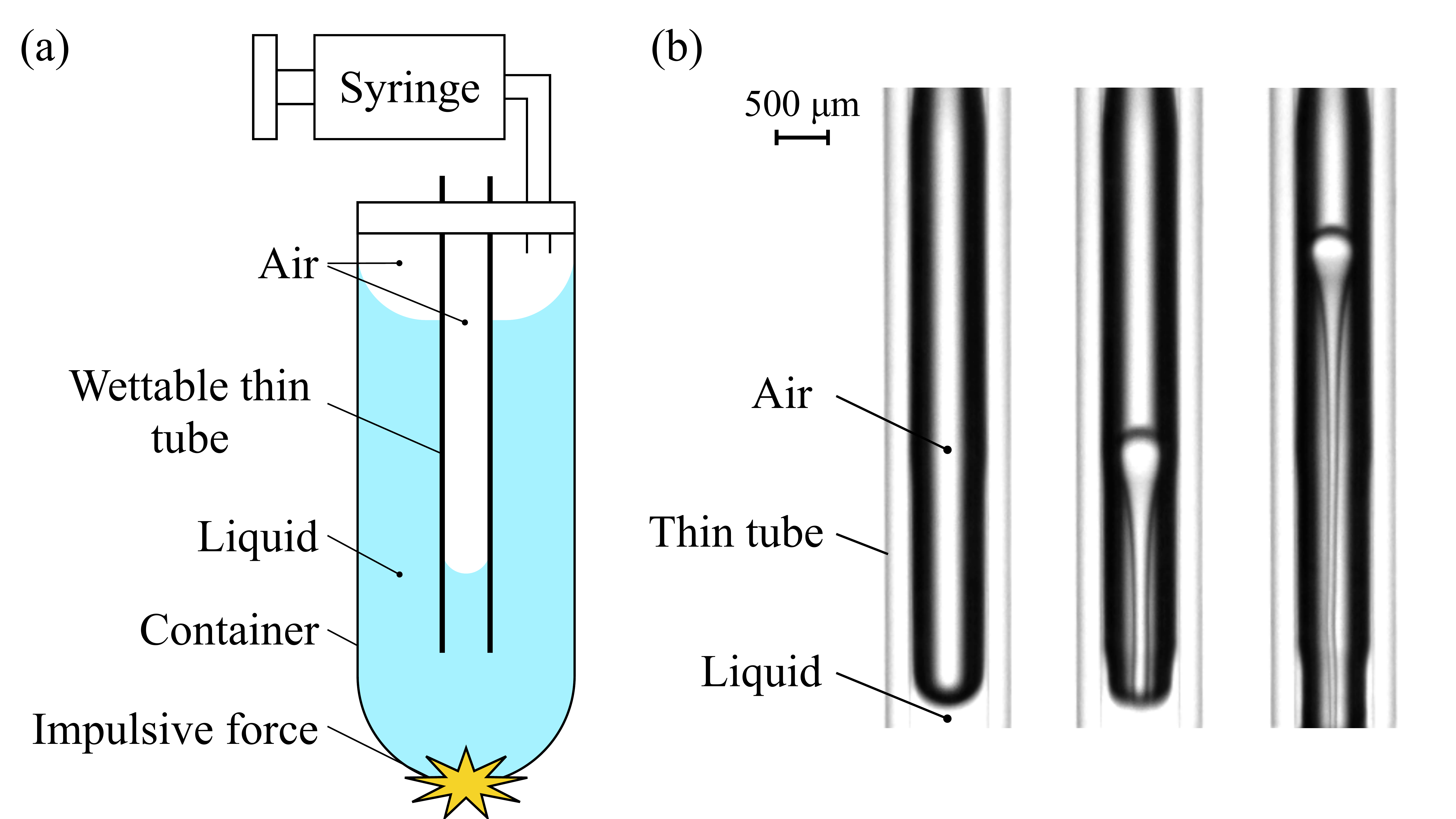}
\caption{\label{dasu-zo}(a) The highly viscous microjet generator.
The liquid level inside the thin tube is kept deeper than that outside the tube by decompressing the air outside the tube.
(b) Snapshots of highly viscous microjet generation (kinematic viscosity $\nu$ = 500 mm$^2$/s).
The images were acquired every $0.37$ ms.}
\end{figure}

Recently, we proposed a simple yet novel technique for generating highly viscous liquid jets \cite{Onuki}.
A schematic illustration of the proposed microjet generator is shown in Fig. \ref{dasu-zo}(a). 
We now briefly explain the idea for generating highly viscous liquid jets \cite{Onuki}, which relies on three tricks:
1) application of an impulsive force on a liquid by applying an impact at the bottom of the liquid-filled container;
2) a thin tube inserted into the liquid, where the liquid level inside the tube is set deeper than that outside the tube;
3) flow-focusing at the gas-liquid interface, which initially has a concave shape.
A focused liquid jet forms at the interface due to the flow-focusing effect \cite{Birkhoff,Milgram,Tagawa_X,Kiyama,Antkowiak,Bergmann}.
Remarkably, these three tricks enabled us to set the velocity of the liquid jet to more than thirty times higher than the initial velocity of the container, resulting in the production of highly viscous liquid jets.
However, the size of the liquid jet was $O(1)$ {\it millimeter}.

In this study, we develop a new device for generating {\it micro}-jets of highly viscous liquids up to 500 mm$^2$/s based on the idea described above (see Fig. \ref{dasu-zo}(b)).
In addition, we analyze the generation process of the microjet by dividing the process into two regimes with different time scales, i.e. {\it Impact time} and {\it Focusing time}.
In Impact time, the liquid accelerates suddenly due to the impulsive force (impact duration $\le O(10^{-4})$ s \cite{Antkowiak}), whereas in Focusing time, the microjet emerges through flow-focusing after the Impact time (focusing duration $\gg O(10^{-4})$ s).

%Highly-viscous microjet generator
\section{BASIC MODEL OF JET GENERATION}
%Physical model
We now explain the physical models for the jet generation in Impact time and Focusing time.

In Impact time, we first consider the model for a setup without the thin tube.
The incompressible Navier-Stokes equation without external forces except gravity is
\begin{equation}
\frac{\partial{\textbf{u}}}{\partial{t}}+\left(\textbf{u}\cdot{\nabla}\right)\textbf{u}=-\frac{1}{\rho}\nabla{p}+\nu\nabla^2\textbf{u}+\textbf{g},
\label{Navier}
\end{equation}
where ${\textbf{u}}$ is the velocity of the liquid, $\rho$ is the density, $p$ is the pressure, $\nu$ is the kinematic viscosity, and $\textbf{g}$ is the acceleration due to gravity.
The liquid is accelerated instantly by an impulsive force \cite{Batchelor,Cooker} from rest to the velocity $U_0$ (referred to as the initial velocity).
During impact, the order of the velocity ${\textbf{u}}$ is given by the initial velocity $U_0$ ($\sim 10^{-1}$ m/s), that of time $t$ is the impact duration $\tau$ ($\sim$ 10$^{-4}$ s \cite{Antkowiak,Kiyama}), and that of $\nabla$ is the inverse of the liquid height $z$ ($\sim 10^{-2}$ m).
We suppose that the kinematic viscosity is much larger than that of water ($\nu \sim 10^{-3}$ m$^2$/s ).
Therefore, the orders of each term in the Navier-Stokes equation are $\partial{\textbf{u}}/\partial{t} = O(10^3)$ m/s$^2$, $\left({\textbf{u}}\cdot\nabla\right){\textbf{u}} = O(1)$ m/s$^2$, $\nu\nabla^2{\textbf{u}} = O(1)$ m/s$^2$, and $\textbf{g} = O(10)$ m/s$^2$.
The inertia term $\left({\textbf{u}}\cdot\nabla\right){\textbf{u}}$, the viscous term $\nu\nabla^2{\textbf{u}}$, and the gravity term ${\textbf{g}}$ are negligible compared to the other terms.
When the device only moves in the vertical direction $z$, temporal integration of Eq. (\ref{Navier}) over the impact duration $\tau$ gives
\begin{eqnarray}
\int_0^{\tau}\frac{\partial\textbf{u}}{\partial{t}}dt&=&-\frac{1}{\rho}\int_0^\tau\nabla{p}dt\nonumber \\
&=&-\frac{1}{\rho}\frac{\partial}{\partial{z}}\int^{\tau}_{0}pdt.
\label{U0}
\end{eqnarray}
The time integral of pressure on the right-hand side of the equation is called a pressure impulse, which is denoted
\begin{equation}
\Pi=\int_0^\tau{p}dt.
\label{Pi}
\end{equation}
When the liquid is accelerated from rest to the initial velocity $U_0$, Eqs. (\ref{U0}) and (\ref{Pi}) yield
\begin{equation}
U_0=-\frac{1}{\rho}\frac{\partial{\Pi}}{\partial{z}}.
\label{model1}
\end{equation}
As shown in Eq. (\ref{model1}), the initial velocity $U_0$ is proportional to the gradient of the pressure impulse in the liquid, which is constant in $z$.
Note that the initial velocity $U_0$ is not affected by the viscosity $\nu$.

\begin{figure}[b]
\includegraphics[scale = 0.13]{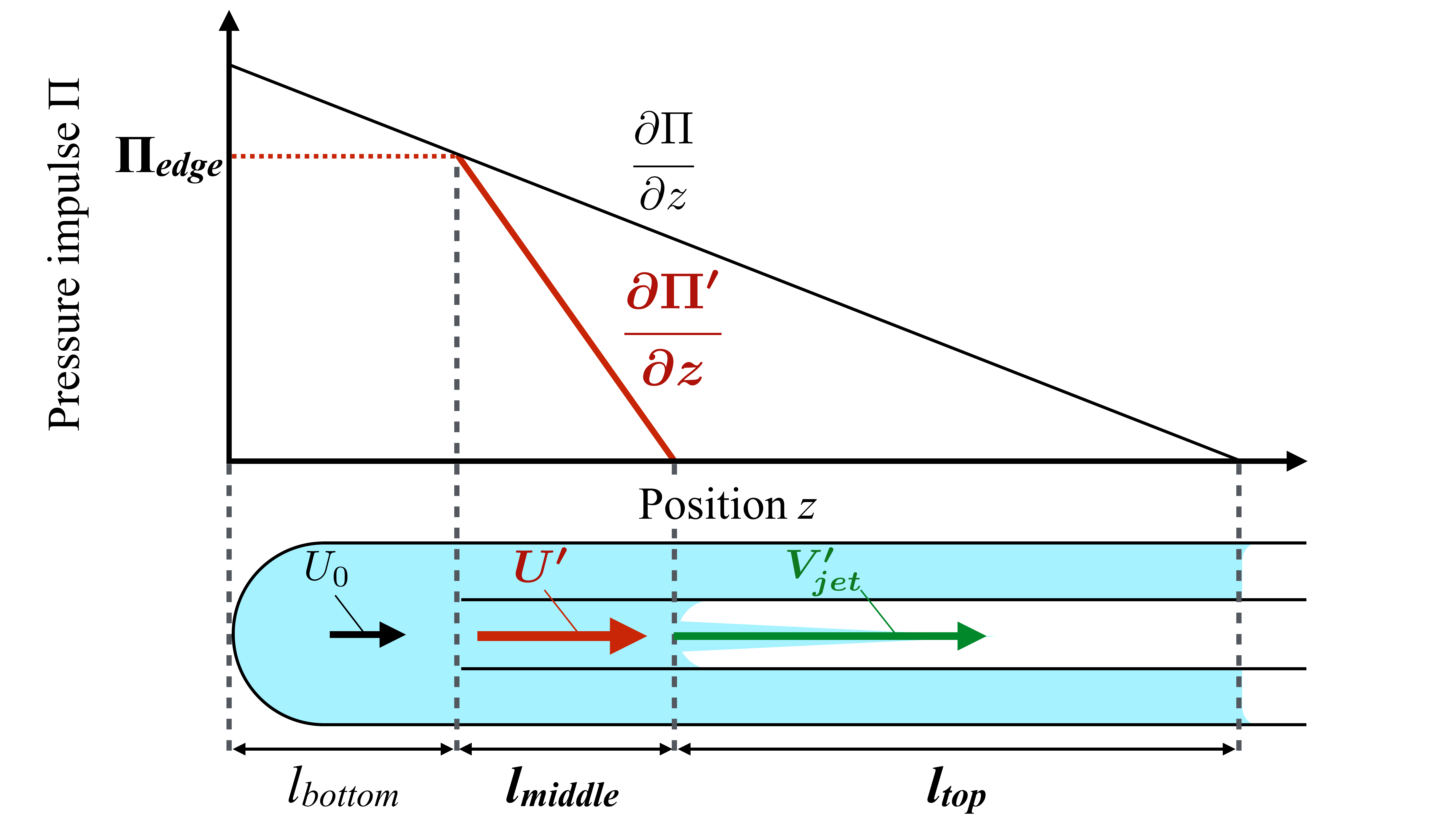}
\vspace{-7mm}
\caption{\label{model}Physical model based on a pressure impulse approach.
The black and red solid lines show the pressure impulse field outside  and inside the thin tube, respectively.}
\end{figure}

\begin{table*}
\caption{\label{parameter}Parameters for experiments and numerical simulations.}
\begin{ruledtabular}
\begin{tabular}{ccccccccccc}
& $\nu$ & $l_{top}$  & $l_{middle}$ & $l_{bottom}$ & $U_0$ & $\theta$\\ \hline
Experiments & 1 - 500 & 39.7 - 61.7 & 1.6 - 10.0 & 16.0 & 0.3 - 0.5 & 28.0 - 35.4\\
Numerical simulations & 1 - 500 & 40.0 - 47.5 & 2.5 - 10.0 & 16.0 & 0.4 & 31.4\\
\end{tabular}
\end{ruledtabular}
\end{table*}

Now we consider the model for a setup with the thin tube in the liquid.
Here we define the parameters shown in Fig. \ref{model}: $l_{bottom}$ is the distance between the bottom of the container and the edge of the thin tube;
$l_{middle}$ is the distance between the edge of the tube and the gas-liquid interface inside the tube; $l_{top}$ is the distance between 
the gas-liquid interface inside the tube and that outside the tube.
When the cross-sectional area of the thin tube is assumed to be much smaller than that of the container, the motion of the liquid outside the tube is described by Eq. (\ref{model1}).
In contrast, the gradient of the pressure impulse inside the tube is significantly larger than that outside the tube since the pressure impulse at the edge of the tube is the same as that outside the tube for a given height (see Fig. \ref{model}).
The pressure impulse at the edge of the tube $\Pi_{edge}$ is calculated using a geometrical relation as
\begin{equation}
\Pi_{edge}=\left(l_{top}+l_{middle}\right)\frac{\partial{\Pi}}{\partial{z}}.
\label{Pi_edge}
\end{equation}
Thus, the gradient of the pressure impulse inside the tube $\partial{\Pi^\prime}/\partial{z}$ is
\begin{eqnarray}
\frac{\partial{\Pi}^\prime}{\partial{z}}&=&\frac{\Pi_{edge}}{l_{middle}}\nonumber\\
&=&\left(\frac{l_{top}}{l_{middle}}+1\right)\frac{\partial{\Pi}}{\partial{z}}.
\label{model2}
\end{eqnarray}
Using Eqs. (\ref{model1}) and (\ref{model2}), the liquid velocity inside the tube $U^\prime$ is
\begin{eqnarray}
U^\prime&=&-\frac{1}{\rho}\frac{\partial{\Pi^\prime}}{\partial{z}}\nonumber\\
&=&\left(\frac{l_{top}}{l_{middle}}+1\right)U_0.
\label{model3}
\end{eqnarray}
Therefore, in Impact time, the liquid velocity inside the tube is controlled by tuning $l_{top}$ and $l_{middle}$.

In Focusing time, the liquid jet is produced by the flow-focusing effect \cite{Birkhoff,Milgram,Tagawa_X,Kiyama,Antkowiak,Bergmann} when the gas-liquid interface reaches velocity $U^\prime$.
The velocity of the microjet $V_{jet}^\prime$ is proportional to the initial velocity at the interface $U^\prime$ \cite{Tagawa_X,Kiyama,Peters}.
Using Eq. (\ref{model3}), the jet velocity inside the tube $V_{jet}^\prime$ is
\begin{eqnarray}
V_{jet}^\prime&=&\beta{U^\prime}\nonumber\\
&=&\beta\left(\frac{l_{top}}{l_{middle}}+1\right)U_0,
\label{model4}
\end{eqnarray}
where $\beta$ is a constant related to the shape of the interface and viscosity of the liquid.
In this paper, $\beta$ is defined to be {\it the increment ratio of the jet velocity}.
The jet velocity $V_{jet}^\prime$ can be varied by changing $l_{top}$ and $l_{middle}$.
Note that $l_{bottom}$ does not affect the jet velocity $V_{jet}^\prime$.

%ExperimentsENumerical simulations
\section{EXPERIMENTS AND NUMERICAL SIMULATIONS}
%Experimental setup
\subsection{Experimental setup}
\begin{figure}[h]
\includegraphics[scale = 0.13]{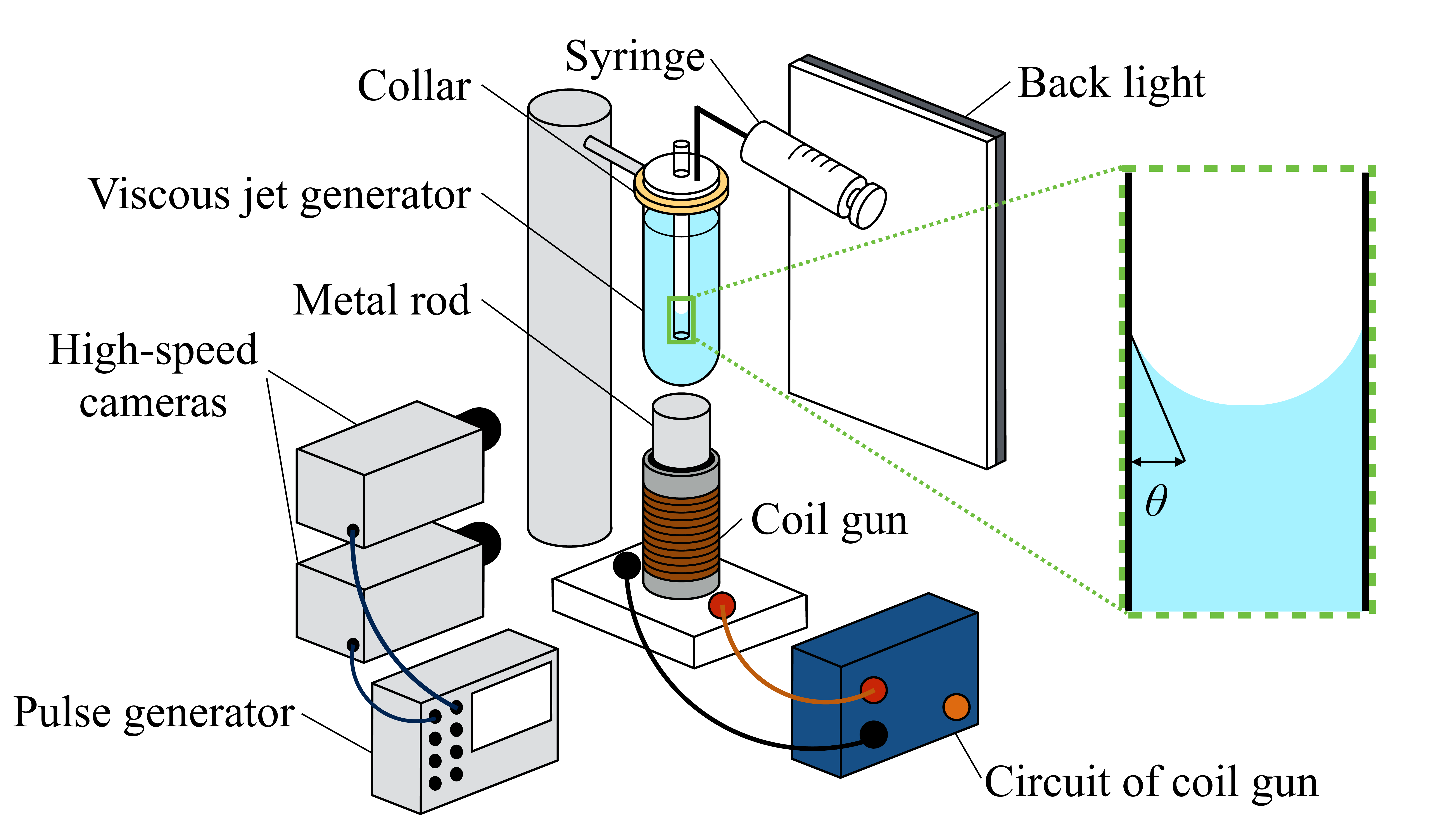}
\caption{\label{apparatus}Sketch of the experimental setup.
$\theta$ is the contact angle between the liquid and the inner wall of the thin tube.}
\end{figure}

The experimental setup is shown in Fig. \ref{apparatus}.
A test tube (A-10, Maruemu, 8.0 mm inner diameter) and silicone oil (KF-96 series, Shin-Etsu Chemical, silicone oil, Sigma Aldrich) are used as the container and liquid, respectively.
A thin-glass tube (FPT-080, Fujiston, 0.5 mm inner diameter, 0.8 mm external diameter) is partially inserted into the liquid in the container.
We control the pressure of the sealed air outside the tube using a syringe.
The liquid level inside the tube is kept deeper than that outside the tube.
To apply the impulsive force, a metal rod is shot (SS400) upward toward the container using a coil gun.
After the rod collides the container, the microjet emerges along the symmetric axis of the interface.
The tip velocity of the microjet $V_{jet}^\prime$ is calculated to be 0.2 ms after the jet passes the three-phase contact point for the first time, and the initial velocity $U_0$ is obtained at 0.8 ms after the jet passing.
We use two high-speed cameras (FASTCAM SA-X, Photron) and a back light (White Led Backlight, Phlox) to capture the motion of the container and the microjet simultaneously.
Both cameras are triggered by a pulse generator (Model 575 Digital Delay/Pulse Generator, BNC).
The frame rate of the camera that photographs the motion of the test tube is 12,500 fps and that of the camera that focuses on the microjet generation is 50,000 fps.
The experimental parameters are summarized in Table \ref{parameter}.
We vary $\nu$, $l_{top}$, $l_{middle}$, and $U_0$, and the experiments are conducted five times for each set of conditions.
The variation in the microjet velocity $V_{jet}^\prime$ owing to the difference in the contact angle $\theta$ is estimated to be less than 8 percent of $V_{jet}^\prime$.

%Numerical setup
\subsection{Numerical setup}
We conducted numerical simulations using a commercial software package employing the finite element method (COMSOL Multiphysics, Keisoku Engineering System Co.).
We separately simulated the liquid acceleration in Impact time and the jet generation in Focusing time under conditions similar to the experimental ones (see Table \ref{parameter}).

\begin{figure}[h]
\includegraphics[scale = 0.13]{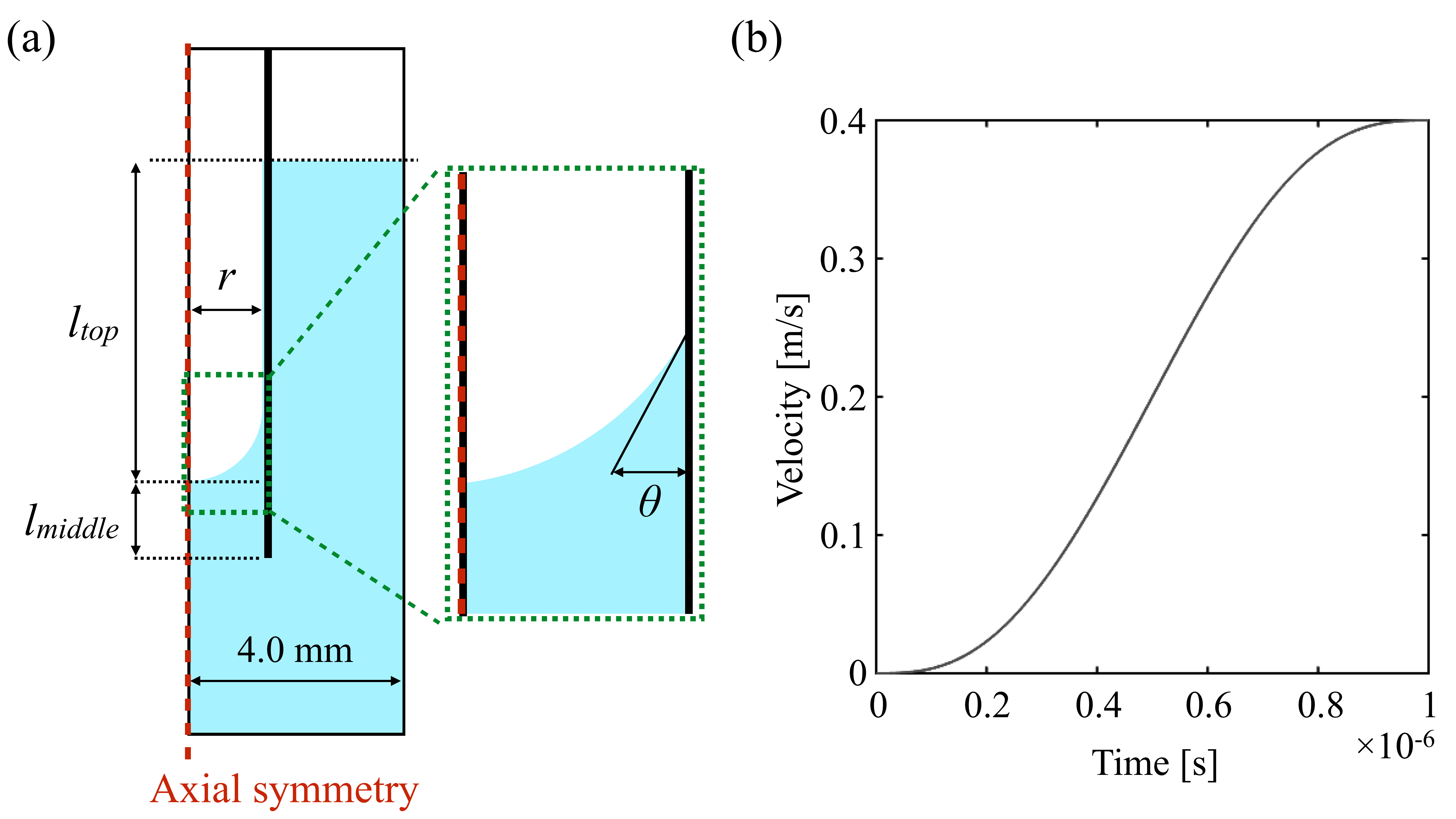}
\caption{\label{geometry1}(a) Geometry for liquid acceleration.
(b) The initial impact applied at the bottom of the geometry.
The velocity at the bottom and the side wall increases until $U_0$ is $1.0\times10^{-6}$ s.
}
\end{figure}

The geometry for calculating the liquid acceleration in Impact time is shown in Fig. \ref{geometry1}(a).
The inner radius of the thin tube $r$ is 0.25 mm.
In Impact time, the distance between the interface inside the tube and that outside the tube is key to increasing the velocity inside the tube.
As an impulsive force, we apply a boundary condition in which the bottom and the side wall accelerate from rest to velocity $U_0$ within $1.0\times10^{-6}$ s (see Fig. \ref{geometry1}(b)).

\begin{figure}[h]
\includegraphics[scale = 0.13]{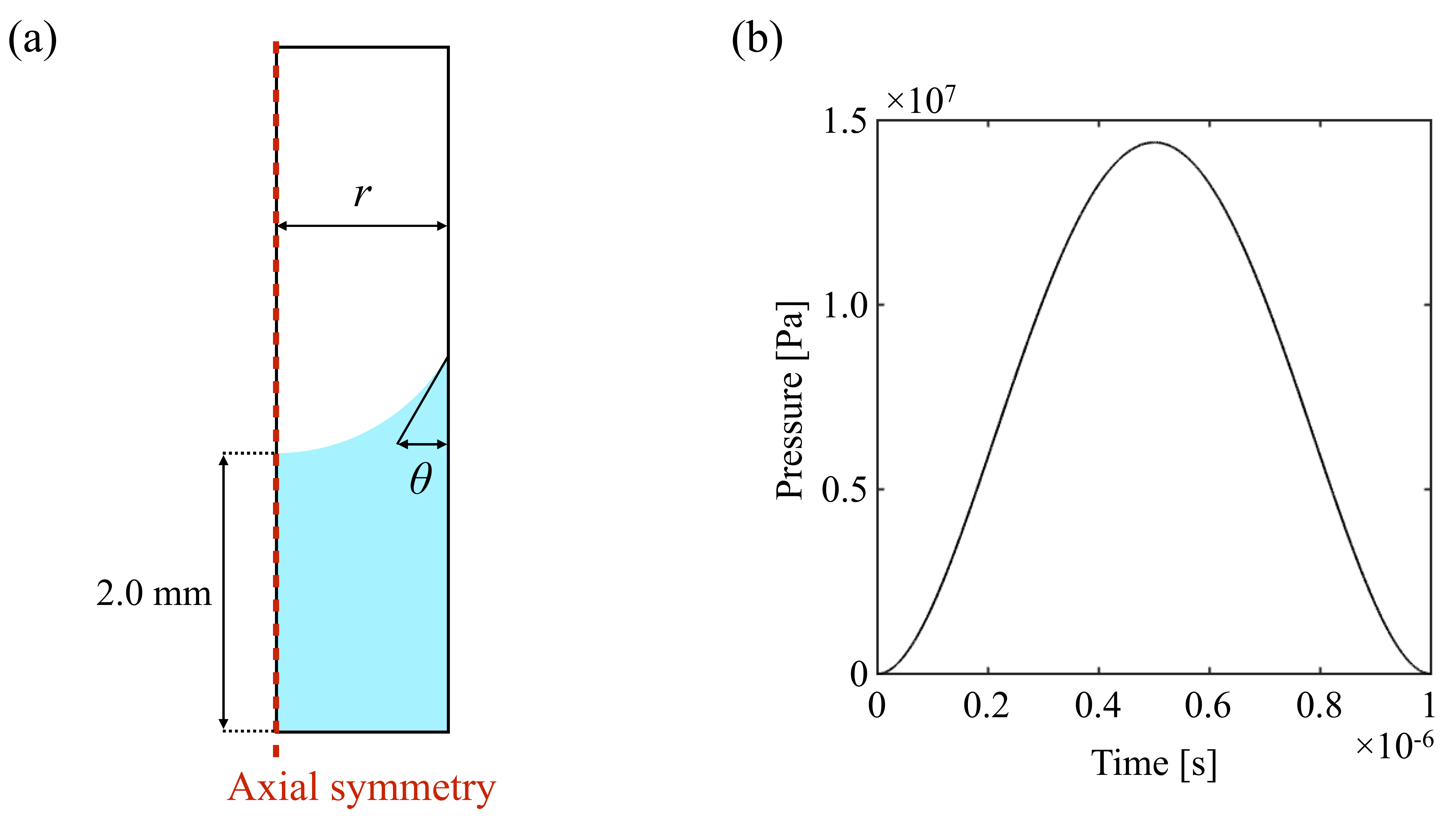}
\caption{\label{geometry2}(a) Geometry for microjet generation.
(b) Boundary condition that enforced at the bottom of the geometry as the sudden acceleration of the liquid.
}
\end{figure}

The geometry for reproducing the microjet generation in Focusing time is shown in Fig. \ref{geometry2}(a).
In Focusing time, the duration of the microjet generation is much longer than the acceleration duration in Impact time.
To reduce the computational load, the simulation for Focusing time is conducted for the liquid inside the tube.
We vary the initial velocity inside the tube $U^\prime$ from 2.0 m/s to 20.0 m/s and the kinematic viscosity $\nu$ from 1 to 500 mm$^2$/s.
As the sudden acceleration of the liquid, the boundary condition that the pressure changes impulsively with in $1.0\times10^{-6}$ s is enforced at the bottom of the geometry (see Fig. \ref{geometry2}(b)).

The governing equations of the simulation are the incompressible Navier-Stokes equation and equation of continuity.
The Level-Set method \cite{level-set,level-set2} is used for tracking the gas-liquid interface.
Level-Set methods define the Level-Set function $\phi$, and $\phi$ = 1 is the liquid phase, $\phi$ = 0 is the gas phase, and 0 $<\phi<$ 1 is the mixed layer of the liquid and the gas.
In the mixed layer, the density $\rho$ and the static viscosity $\mu$ are calculated as
\begin{eqnarray}
\rho &=& \rho_{l}\phi+\rho_{g}(1-\phi)\label{rho_eq}\\
\mu &=& \mu_{l}\phi+\mu_{g}(1-\phi).\label{mu_eq}
\end{eqnarray}
The suffixes $l$ and $g$ in Eqs. (\ref{rho_eq}) and (\ref{mu_eq}) indicate the liquid and gas phases, respectively.
In this paper, we define $\phi$ = 0.5 to be the gas-liquid interface.
Using the Level-Set function, the movement of the interface is expressed as
\begin{equation}
\frac{\partial\phi}{\partial t}+{\bf{u}}\cdot\nabla\phi=\gamma\nabla\cdot\left(\epsilon\nabla\phi-\phi\left(1-\phi\right)\frac{\nabla\phi}{|\nabla \phi|}\right),
\label{level-set}
\end{equation}
where $\gamma$ is the parameter that determines the amount of reinitialization.
The $\epsilon$ is the interface thickness defined as half of the characteristic mesh size near the interface.
In this simulation, we use a triangle mesh, the maximum size of which is 0.05 mm.

\begin{figure}[!b]
\includegraphics[scale = 0.22]{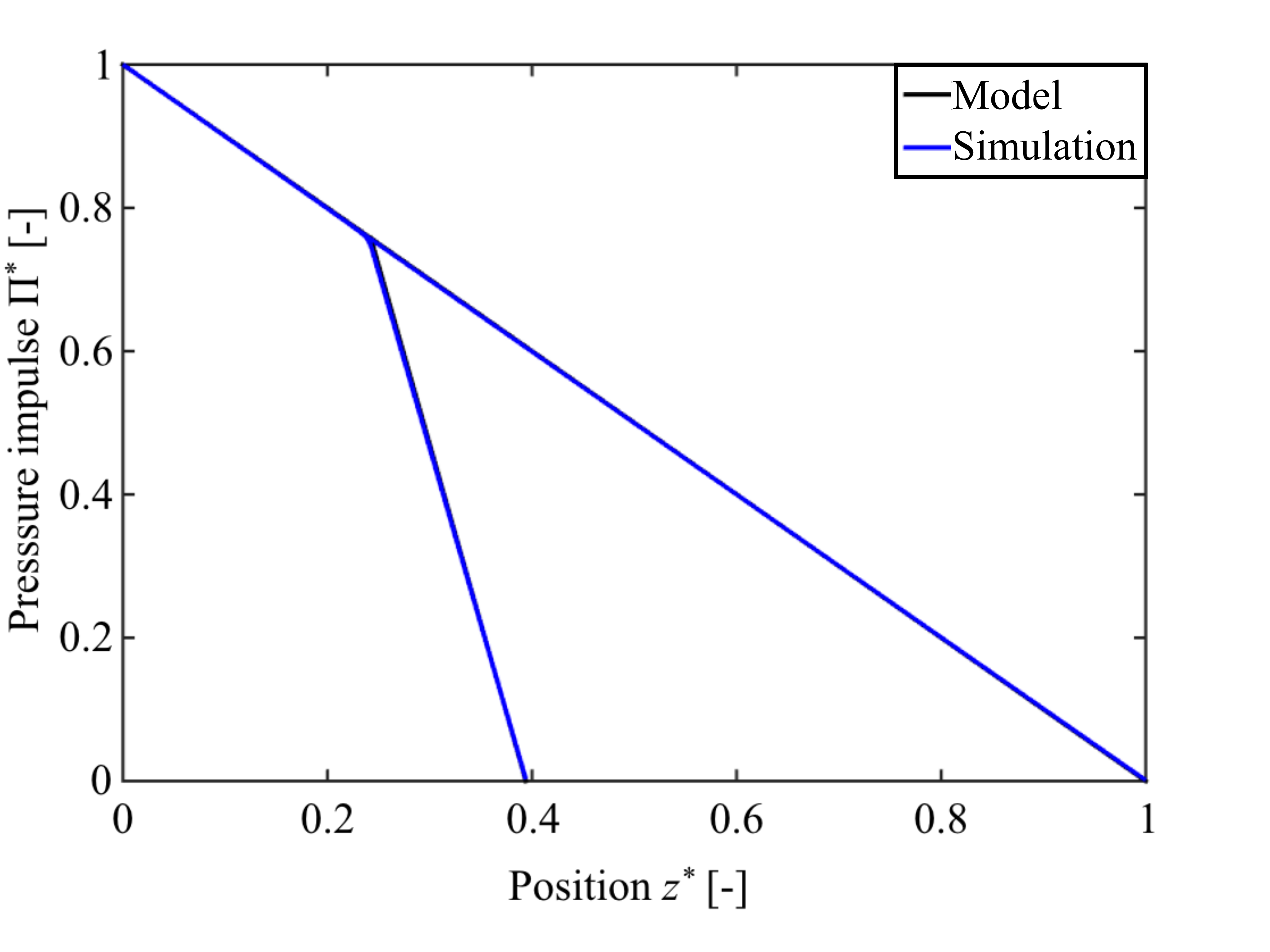}
\vspace{-1em}
\caption{\label{contour}Relation between the pressure impulse $\Pi^*$ and the vertical position from the bottom of the container $z^*$ just after impact ($t=1.0\times10^{-6}$ s).
$l_{top}$ is 40.0 mm, $l_{middle}$ is 10.0 mm, and $l_{bottom}$ is 16.0 mm.
The kinematic viscosity $\nu$ is 1 mm$^2$/s.}
\end{figure}

\begin{table}[!b]
\caption{\label{pi_u_l}Relation between $(l_{top}/l_{middle}+1)$, the gradient of the pressure impulse and the velocity inside the tube.
$|\partial{\Pi^\prime{^*}}/\partial{z^*}|$ is the gradient of the pressure impulse inside the tube divided by that outside the tube, and $U^*$ is the normalized velocity inside the tube divided by that outside the tube $U_0$.
}
\begin{ruledtabular}
\begin{tabular}{cccc}
 $(l_{top}/l_{middle}+1)$ [-]& $\nu$ [mm$^2$/s]&$|\partial{\Pi^\prime{^*}}/\partial{z^*}|$ [-]& $U^*$ [-]\\ \hline
& 1 & 4.9 & 4.9\\
5& 100 & 4.9 & 4.9\\
& 500 & 4.9 & 4.9\\ \hline
& 1 & 9.5 & 9.6 \\
10& 100 & 9.5 & 9.4 \\
& 500 & 9.5 & 9.4 \\ \hline
& 1 & 13.8 & 13.4 \\
15& 100 & 13.8 & 13.4 \\
& 500 & 13.8 & 13.4 \\ \hline
& 1 & 18.1 & 17.3 \\
20& 100 & 18.0 & 17.3 \\
& 500 & 18.0 & 17.3 \\
\end{tabular}
\end{ruledtabular}
\end{table}

%Results
\section{RESULTS}
\subsection{Impact time: acceleration of the liquid}
We show the simulated pressure impulse $\Pi^*$ normalized by that at the bottom of the container as a function of the non-dimensional vertical position from the bottom $z^*$ normalized by that of the interface outside the tube (= $ l_{bottom}+l_{middle}+l_{top}$) in Fig. \ref{contour} as the blue solid line.
The gradient of the pressure impulse inside the tube $\partial{\Pi^\prime{^*}}/\partial{z^*}$ is much larger than that outside the tube.
The model calculated from Eq.({\ref{model2}}) is shown in Fig. \ref{contour} as a black solid line.
The model is in good agreement with the numerical result.
The gradient of the pressure impulse and the velocity inside the tube for each set of conditions are shown in Table \ref{pi_u_l}.
Here, $U^\prime{^*}$ is the velocity inside the tube normalized by that outside the tube $U_0$.
The velocity inside the tube $U^\prime{^*}$ increases up to 17.3 times that outside the tube.
The gradient of the pressure impulse $\partial{\Pi^\prime{^*}}/\partial{z^*}$ and the velocity $U^\prime{^*}$ are in reasonable agreement with $(l_{top}/l_{middle}+1)$ for all the conditions evaluated, as expected from Eqs. (\ref{model2}) and (\ref{model3}).
Remarkably, the gradient of the pressure impulse $\partial{\Pi^\prime{^*}}/\partial{z^*}$ and the velocity $U^\prime{^*}$ in each $(l_{top}/l_{middle}+1)$ are almost constant even for the highly-viscous liquid.
The viscous effect on the sudden acceleration is thus negligible in Impact time.

\begin{figure}[!b]
{\includegraphics[scale = 0.22]{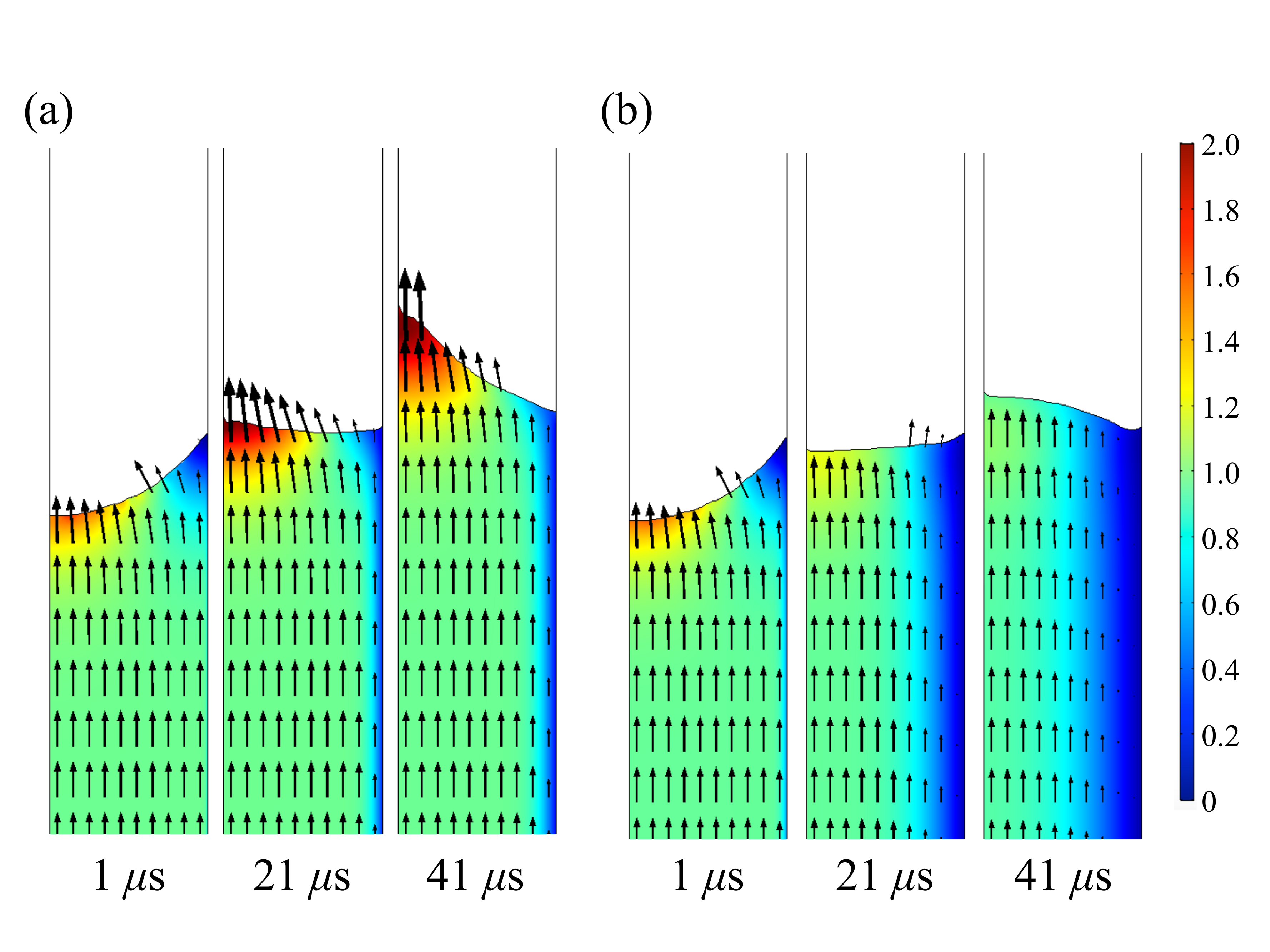}}
\vspace{-1em}
\caption{\label{vector}Vector fields in the liquid during the deformation of the gas-liquid interface.
Results with (a) small viscosity $\nu = 10$ mm$^2$/s and (b) large viscosity $\nu = 100$ mm$^2$/s.
The initial velocity $U^\prime$ = 4.0 m/s is applied to both conditions.
The left-hand image for each condition corresponds to the moment that the sudden acceleration is finished.
}
\end{figure}

\begin{figure*}[t]
{\includegraphics[scale = 0.22]{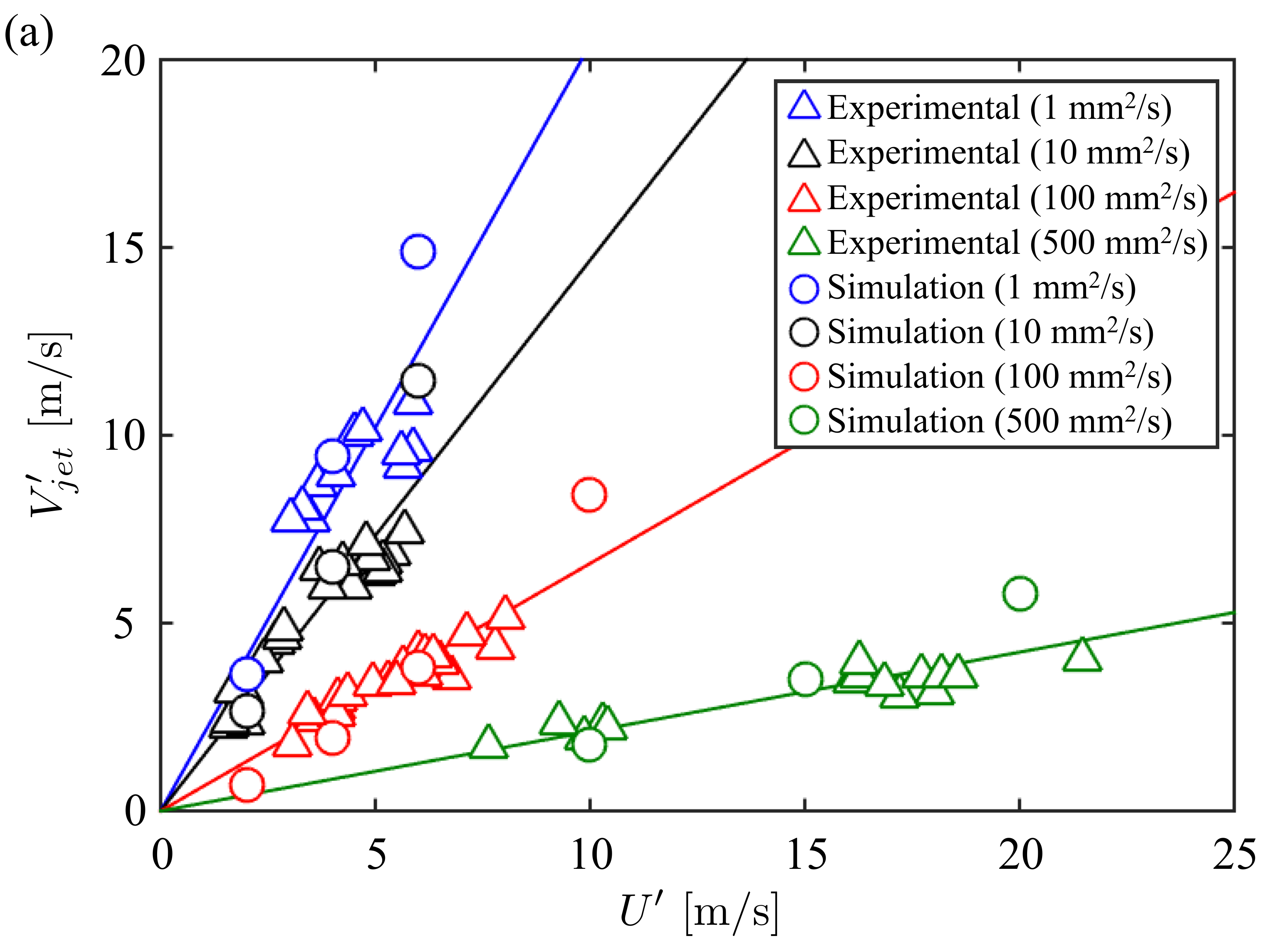}}
\hspace{2mm}
\includegraphics[scale = 0.22]{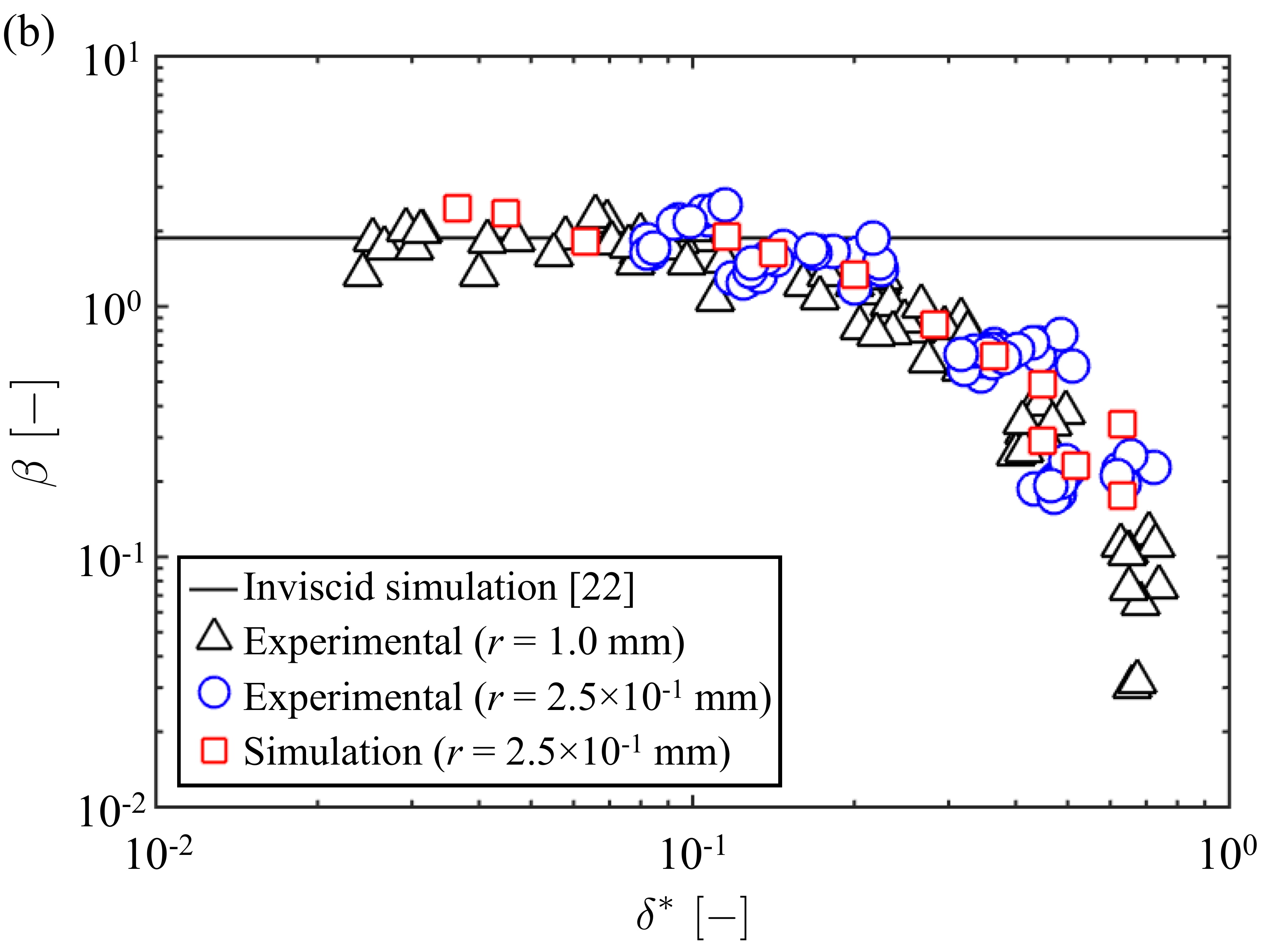}
\caption{\label{result}
(a) Relation between the velocity of the microjet $V_{jet}^\prime$ and the initial velocity inside the tube $U^\prime$.
The triangles are the experimental data and the circles are the simulations.
Blue, black, red, green correspond to results with 1 mm$^2$/s, 10 mm$^2$/s, 100 mm$^2$/s and 500 mm$^2$/s, respectively.
(b) Increment ratio of the jet velocity $\beta(=V_j^\prime/U^\prime,\ {\rm{see\ Eq. (\ref{model4})}})$ vs. normalized boundary layer thickness $\delta^*$.
The circles and squares are the experimental and simulation results, respectively.
The triangles are experimental data obtained using a thin tube with a large inner radius $r = 1$ mm ($\nu$ = 1 - 1,000 mm$^2$/s) \citep{Onuki}.
}
\end{figure*}

\subsection{Focusing time: microjet generation}
The velocity fields during flow-focusing are shown in Fig. \ref{vector}.
Here, the velocity of each point is normalized by the initial velocity applied at the bottom of the geometry $U^\prime$.
The velocity field just after the impact ($t=1.0\ \mu$s) is almost constant regardless of the viscosity $\nu$, i.e. the viscous effect does not appear just after the impact.
This trend agrees with the simulation results shown in Section IV(A).
For a low-viscosity liquid (see Fig. \ref{vector}(a)), after $t=1\ \mu$s, the velocity of the center of the interface increases up to twice the initial velocity $U^\prime$ thanks to the flow-focusing effect.
On the other hand, for a high-viscosity liquid (see Fig. \ref{vector}(b)), the increment of the velocity due to flow-focusing does not occur.
This indicates that the liquid viscosity affects the deformation of the interface during flow-focusing.
Here, we further discuss the viscous effect on microjet velocity.

We show the relation between the microjet velocity $V_{jet}^\prime$ and the initial velocity inside the tube $U^\prime$, both of which are obtained from the experiments and the numerical simulations (Fig. \ref{result}(a)).
Note that the initial velocity $U^\prime$ in the experiments is calculated from Eq. (\ref{model3}).
For each viscosity $\nu$, the simulations are in agreement with the experiments, which demonstrates that the simulations can fairly well reproduce the high-viscosity microjets generated in the experiments.

The model (see Eq. (\ref{model4})) for each viscosity $\nu$ is shown in Fig. \ref{result}(a) as the solid line.
Note that $\beta$ in Eq. (\ref{model4}) is fitted to both the experimental and simulation results for each viscosity $\nu$.
The increment rate of the jet velocity $\beta$ decreases with increasing viscosity $\nu$.
In other words, in Focusing time, the viscous effect appears strongly for liquid jets with high viscosity.

To clarify the viscous effect on $\beta$, we consider the development of the boundary layer during the deformation of the interface.
It is known that the thickness of the boundary layer $\delta$ is estimated as $\delta = \sqrt{\nu{t}}$.
The focusing time scale $t_f$ is estimated as \citep{Tagawa_X}
\begin{equation}
t_f \sim \frac{2r}{U^\prime}.
\label{tf}
\end{equation}
The normalized thickness of the boundary layer $\delta^*$ is defined as
\begin{eqnarray}
\delta^*&=& \frac{\left.\delta \right|_{t = t_f}}{r}\nonumber\\
&=&\sqrt{\frac{2}{Re}}.
\label{delta}
\end{eqnarray}
Here, Reynolds number $Re$ is calculated as $Re = U^\prime{r}/\nu$.
The normalized thickness of the boundary layer $\delta^*$ is inversely proportional to $\sqrt{Re}$.
Because the boundary layer intercepts flow-focusing, the increment ratio of the jet velocity $\beta$ would decrease with increasing $\delta^*$.
The relation between the increment ratio of the jet velocity $\beta$ and the normalized thickness of the boundary layer $\delta^*$ is shown in Fig. \ref{result}(b).
Note that the increment ratio $\beta$ is calculated for each result.
The experimental results obtained using a tube with a large inner radius $r = 1$ mm ($\nu$ = 1 - 1,000 mm$^2$/s) \citep{Onuki} are also shown in Fig. \ref{result}(b).
For various values of inner radius $r$ and viscosity $\nu$, the trend of the increment ratio $\beta$ is described by using the normalized thickness of boundary layer $\delta^*$, namely Reynolds number $Re$.
Here, we show the $\beta$ obtained from inviscid simulation \citep{Peters} in Fig. \ref{result}(b) as a black solid line.
The inviscid simulation \citep{Peters} agrees well with our experiments and simulations carried out with small $\delta^*\lesssim10^{-1}$ ($Re\gtrsim200$).
On the other hand, for $\delta^*\gtrsim10^{-1}$ ($Re\lesssim200$), the $\beta$ decreases with increasing $\delta^*$, which indicates that in this range ($Re\lesssim200$) the growth of boundary layer affects the microjet generation in Focusing time.

%Conclusion
\section{CONCLUSION}
We proposed a new device for generating highly viscous microjets based on some simple tricks: 1) application of an impulse on the bottom of the container;
2) a submerged thin tube, for which the liquid level inside the tube is set deeper than that outside the tube;
3) preparation of a concave interface for the generation of flow-focusing.
We conducted the microjet generation experiments and found that our device is able to produce highly viscous liquid microjets (up to 500 mm$^2$/s).
To determine the generation mechanism, we divided the process into two parts:
Impact time is the period during which the liquid is suddenly accelerated due to the impulsive force and Focusing time is the period during which the liquid emerges as the microjet due to flow-focusing.
In Impact time, we focused on the pressure impulse just after sudden acceleration.
We developed a physical model and validated it by comparing it with the numerical simulations.
We found that the model is able to describe the simulated pressure impulse fields, which indicates that the basic mechanism of the device is well understood.
In Impact time, the effect of viscosity is negligible.

In contrast, higher viscosities decrease the velocity of the microjet in Focusing time.
To reveal the mechanism, we consider the development of the boundary layer since the viscosity interrupts flow-focusing.
Remarkably, the decrement of the jet velocity is found to be described by using Reynolds number $Re$.

It is worth mentioning that the device is capable of ejecting the jets in multiple directions.
This enables us to print viscous liquids to an object with irregular surfaces.
Thanks to its simple structure, we can easily produce a hand-sized device at low cost.

%Acknowledgements
\begin{acknowledgements}
This work was supported by JSPS KAKENHI Grant Number 26709007,
17H01246, and 17J06711.
\end{acknowledgements}

\nocite{*}

\bibliography{apssamp}% Produces the bibliography via BibTeX.

\end{document}